\documentstyle{lamuphys}
\makeatletter
\let\chapter\hid@chapter
\makeatother
\newcommand{\be}{\begin{equation}}
\newcommand{\ee}{\end{equation}}
\newcommand{\bea}{\begin{eqnarray}}
\newcommand{\eea}{\end{eqnarray}}

\begin{document}
\pagenumbering{arabic}
\title{Towards a Full Quantum Theory of Black Holes}

\author{Claus\,Kiefer}

\institute{Fakult\"at f\"ur Physik, Universit\"at Freiburg,
Hermann-Herder-Stra\ss e 3, D-79104 Freiburg, Germany}

\maketitle

\begin{abstract}
This review gives an introduction to various attempts to
understand the quantum nature of black holes. The first part
focuses on thermodynamics of black holes, Hawking radiation,
and the interpretation of entropy. The second part is devoted
to the detailed treatment of black holes within canonical
quantum gravity. The last part adds a brief discussion 
of black holes in string theory and quantum cosmology.
\end{abstract}

\small
\noindent Report
     Freiburg THEP-97/24; to appear in {\em Black Holes: Theory and
     Observation}, edited by F.W. Hehl, C. Kiefer, and R. Metzler
     (Springer, Berlin, 1998).
\normalsize

\section{Introduction and Summary}
It is of fundamental importance to obtain a full quantum description
of black holes. The reasons are of a technical, conceptual, and
observational nature. Technical, because it provides a highly
nontrivial application of quantum gravitational equations
in the full, non-perturbative, regime. One of the main open issues thereby
is what substitutes the classical singularities in quantum theory.
 Conceptual, because
the present status of semiclassical approaches leads to problems
such as the information loss problem, which can be 
satisfactorily dealt with only in the full theory.
Observational, because apart from potential cosmological
data this is probably the only window to directly test a quantum
theory of gravity. 

This goal has not yet been reached, since a consistent theory of
quantum gravity has not yet been constructed.
Many quantum aspects of black holes, however, have been
understood in the last 25 years, which could lead the way to a
full understanding. This review article is intended to give
a pedagogical introduction to results which have been obtained
in the framework of present approaches towards quantum gravity.

In Sect.~2, I shall review the key issues which lead to
the conclusion that black holes are quantum objects. The issues are
thermodynamics of black holes, Hawking radiation, and the
interpretation of black hole entropy. Since many of these topics
are discussed at great length by other lecturers, in particular by
't~Hooft, Israel, Neugebauer, and Wipf, I shall present only   
those issues which I consider to be of particular relevance.

Sect.~3 presents one approach towards a theory of quantum
gravity in some detail -- the canonical quantisation of general relativity.
This approach by itself most likely leads to
an effective theory only, but it is
the most straightforward approach available and offers by itself
interesting insights into possible quantum aspects of black holes.
The issues addressed cover both applications of the ``full"
theory (such as a wave function for the eternal Reissner-Nordstr\"om hole)
and the semiclassical expansion
(such as the description of Hawking radiation and black hole entropy in
the context of the Wheeler-DeWitt equation).

Sect.~4, finally, gives a brief introduction to superstrings and the
issue of black hole entropy as being obtained from counting
the states of D-branes. I shall also offer some speculations
about the role of black holes in quantum cosmology. 
  
\section{Why Black Holes are Quantum Objects}

\subsection{Thermodynamics of Black Holes}
In the beginning of the seventies, a surprising analogy
was discovered between black holes and thermodynamical systems
in the framework of general relativity,
see the lectures by Israel and Neugebauer in this volume.
(Other reviews are, e.g., Bekenstein (1980), Wald (1994, 1997a),
and Kiefer (1997a)). This analogy is summarised in
Table~1 (with an obvious notation):

\begin{table}[htb]
\caption[ ]{The laws of black hole mechanics}
\begin{flushleft}
\renewcommand{\arraystretch}{1.2}  
\begin{tabular}{l|l|l}
{Law} & {Thermodynamics} &  Stationary Black Hole\\ \hline
{} & {} & {} \\
Zeroth & \ $T$ constant on a body &\ surface gravity
                                    $\kappa$ constant on the \\
{} &\ in thermal equilibrium & \ horizon of a black hole\\
{} & {} & {} \\
First & $\ \D E=T\D S-p\D V+\mu dN$ &
 $\ \D(mc^2)=\frac{\kappa c^2}{8\pi G}\D A
+\Omega \D J-\phi \D q$ \\ 
{} & {} & {} \\
Second &\ $\D S\geq 0$ & $\ \D A\geq 0$ \\
{} & {} & {} \\
Third &\ $T=0$ cannot be reached &\ $\kappa=0$ cannot be reached\\
\end{tabular}
\renewcommand{\arraystretch}{1}
\end{flushleft}\end{table}

In the following I shall mostly deal with nonrotating holes
($J=0$), but often keep a nonvanishing charge $q$.
This is not realistic from an astrophysical point of view,
but provides an interesting nontrivial example which mimics
in many examples the relevant case of rotating holes. 

Some comments are appropriate for the Third Law, because this
will also be relevant for Sect.~3. In ordinary thermodynamics,
there exist various inequivalent formulations of this law.
One version frequently used was introduced by Planck in 1911: 
The entropy $S$ goes to zero (or a material-dependent constant) as
the temperature $T$ goes to zero. From this (and some mild
assumptions) follows a weaker version: $T=0$ cannot be reached
in a finite number of steps, see e.g. Wilks (1961) for details.
It is {\em this} version of the Third Law that was proven
by Israel (1986) for black holes and that is stated in 
Table~1. (In the proof the validity of the weak energy condition
for matter in a neighbourhood of the apparent horizon was used.)

\vskip 2mm
\small

$S\to 0$ as $T\to 0$ is very helpful in thermodynamics,
since it allows one to determine the entropy from measurements
of specific heats, $C$. It follows from Planck's version of
the Third Law that $C\to 0$ as $T\to 0$, but not vice versa
(as is sometimes erroneously stated). Planck's version is
not always fulfilled; it is violated, for example, for glasses
(which have a higher disorder than the corresponding cristalline
state). Other examples include the molecule $CH_3D$ (Straumann~1986)
or a gas confined to a circular string at zero temperature
(Wald~1997b). From the point of view of quantum statistics
it is clear that Planck's version holds if there is a unique
non-degenerate ground state at $T=0$. This is violated in these
examples.

\vskip 2mm
\normalsize

The above analogy between black hole mechanics and
ordinary thermodynamics holds in a much more general framework
than general relativity,
see Iyer and Wald (1994, 1995), and Wald (1997a).
 If one only assumes that the field
equations follow from a diffeomorphism covariant Lagrangian, $L$,
the First Law holds (whether a generalisation
of the area theorem holds is not clear). 

\vskip 2mm
\small

The term\footnote{From now on we set $c=1$.}
 $\kappa \D A/8\pi G$ occurring in the First Law 
is replaced by
\be \D\int_{\cal C}\mbox{Q}=\frac{\kappa}{2\pi}\D{\cal S}, \quad
    \mbox{with } {\cal S}\equiv -2\pi\int_{\cal C}
     \frac{\delta L}{\delta R_{abcd}}n_{ab}n_{cd} \enspace, \ee
where $\mbox{Q}$ is the Noether charge 2-form associated with the
Killing field $\xi$ normal to the horizon, where the presence
of a bifurcate Killing horizon is assumed (${\cal C}$ is the
bifurcation surface); $n_{ab}$ denotes the binormal to
${\cal C}$ ($\nabla_a\xi_b=\kappa n_{ab}$). For the special case
of general relativity, $L=R\sqrt{-g}/16\pi G$, the corresponding
expression in Table~1 is recovered. If one, on the other hand,
assumes beforehand that $S\propto A$,
the Einstein field equations must hold (Jacobson~1995).

\vskip 2mm
\normalsize

For generalisations of the laws of black hole mechanics to
cases where nonabelian matter fields are present I refer
to Heusler (1996), and the references therein.

For completeness I want to mention another, different, analogy between
black holes and statistical mechanics: 
Choptuik (1993) discovered through numerical studies
 that if a spherical wave packet of a massless
scalar field collapses, there exists a critical parameter
(characterising the strength of the ensuing gravitational
self-interaction of the field)
above which no black hole forms. In the vicinity of this
critical parameter there is a universal relation for
the black hole mass like in the vicinity of a critical point
in statistical mechanics. 

\subsection{Hawking Radiation}

The analogies between ordinary thermodynamics and black hole
mechanics, summarised in Table~1, were first regarded as purely formal,
since classically a black hole cannot radiate
(it behaves like an ideal absorber). Can quantum theory
change this conclusion? One could imagine that $T_{BH}\propto\hbar$
and $S_{BH}\propto\hbar^{-1}$; in fact, from dimensional
arguments one recognises that to achieve $T_{BH}\neq0$ one would
have $T_{BH}\propto \hbar\kappa/k_B$ and $S_{BH}\propto k_BA/G\hbar$,
since no other fundamental constants are at one's disposal
(at least within standard physics). 

Using quantum field theory on a curved background
spacetime, Hawking (1975) was able to show that black holes
{\em do} in fact radiate and have a finite entropy. The temperature is
\be T_{BH}=\frac{\hbar\kappa}{2\pi k_B}\ , \ee
and the entropy therefore from the First Law
\be S_{BH}=\frac{k_BA}{4G\hbar}\enspace . \ee
This is a very general result, since no use of particular
gravitational field equations was made.

\vskip 2mm
\small

This is the reason why black hole thermodynamics seems to hold
in a much wider framework, see (1). One there has
the formal expression $S_{BH}=
k_B{\cal S}/\hbar$, which would thus give a general local geometric
notion of black hole entropy. However, no quantum field theoretical
calculation has been made to justify this interpretation.

\vskip 2mm
\normalsize

For later convenience I give the explicit expressions for a
Reissner-Nordstr\"om black hole (a charged spherically symmetric
black hole),
\be T_{BH}=\frac{\hbar}{8\pi k_BGm}\left(1-\frac{q^4}{R_0^4}
    \right)\enspace , \ee
where
\be R_0=Gm+\sqrt{(Gm)^2-q^2} \ee
is the radius of the event horizon. The entropy is
\be S_{BH}=\frac{k_B}{G\hbar}\pi R_0^2\enspace .\ee
An extremal hole is defined by $\vert q\vert=Gm$; its temperature
thus vanishes, while its entropy is nonvanishing
 (and not a constant).
It thus seems as if Planck's version of the Third Law were violated,
but the situation for extremal holes is more subtle, as will
be discussed in Sect.~3. Holes with $\vert q\vert>Gm$ exhibit
a naked singularity and are therefore generally excluded
from consideration, although their role within quantum gravity
is unclear.

How can one interpret Hawking radiation? The central point is that
the notion of vacuum (and therefore also the notion of particles)
loses its invariant meaning in the presence of a dynamical
background. Incoming modes of the quantum field are redshifted
while propagating through the collapsing geometry, which is why the
quantum state of the outgoing modes is different. If the
initial state is a vacuum state, the outgoing state contains
``particles". The redshift is especially high near the
horizon, where the modes spend a long time before escaping
to infinity. This is the reason why Hawking radiation is present
very long after the collapse is finished for a comoving
observer, contrary to what one
would naively expect. The presence of the horizon is also
responsible for the thermal nature of the radiation, since
no particular information about the details of the collapse
can enter. It turns out that the vacuum expectation value
of the energy-momentum tensor of the quantum field is negative
near the horizon,
corresponding to a flux of negative
energy {\em into} the hole (this is the basis for the pictorial interpretation
of the Hawking effect, where one partner of a pair of virtual
particles can fall into the hole, thus enabling the other partner
to become real and escape to infinity, where it can be observed as
Hawking radiation). For details of this scenario, I refer to e.g.
Wipf (this volume), 't~Hooft (1996, and this volume),
 Birrell and Davies (1982), Wald (1994),
and the references therein. The negativity of this expectation value
is, like the Casimir effect, a genuine quantum feature.

This negative energy flux leads to a decrease of the black hole mass
and is equal to the positive flux of the Hawking radiation
at infinity. From a simple application of Stefan-Boltzmann's law,
one can heuristically estimate that the time $t(m_P)$ for the hole
to lower its mass to roughly the Planck mass $m_P\equiv\sqrt{\hbar/G}$ is
$t(m_P)\propto m_0^3$, where $m_0$ is the initial mass of the hole.
After this stage is reached, the semiclassical calculations
used by Hawking (1975) are expected to break down. It is one of the
most interesting open features of a full quantum gravity to
provide a detailed understanding of this final phase.

\vskip 2mm
\small

How can one observe the Hawking effect? It is easy to estimate that
for an initial mass of about one solar mass, $m_0\approx m_{\sun}$,
$t(m_P)\approx 10^{65}\mbox{yrs}$, which is much longer than the
age of the Universe. Before this time the radiation is much too weak
to be noticeable.
 The effect can thus {\em not} be observed
for black holes originating from stellar collapse. Only if
primordial black holes were left over from the Big Bang,
would there be a hope of observation (if the initial mass of
the hole is $m_0\approx 10^{15}\mbox{g}$,\footnote{The size of such
a hole would be only about $10^{-13}\mbox{cm}$!}
 the final stages
of the primordial hole would occur ``today"). The amount of primordial
holes is strongly constrained by the smoothness of the Big Bang,
see Sect.~4. It is thus not clear whether this effect is observable
at all. Bousso and Hawking (1997) have investigated pair creation
of black holes during an inflationary phase in the early Universe.
By applying the no-boundary proposal of Hartle and Hawking (1983),
they estimated that no significant number of neutral holes
having sufficient initial mass survive inflation.

If ``hot" black holes were around, they would contribute to the
observed $\gamma$-ray background. Before the final evaporation
(about which nothing is known), the spectrum should
according to (2) be thermal. Since this is not true for
the $\gamma$-ray background, one finds from observations that
the number of primordial holes must be less than about $10^4$
per $\mbox{pc}^3$ (Page and Hawking~1976). Wright (1996) estimated
from the anisotropy component of the $\gamma$-ray background in the 
halo of the Milky Way an upper limit of $0.4$ explosions
of primordial holes per $\mbox{pc}^3$ and year. 

It may also be possible that the existence of primordial black holes
can be inferred from the variation of quasar luminosities
(Hawkins~1993), although this is at present a contentious issue.

\vskip 2mm
\normalsize

It must be mentioned that there exists an effect analogous to
the Hawking effect in Minkowski space, discovered by Unruh (1976).
An observer with uniform acceleration $a$ observes thermal
radiation in the Minkowski vacuum state with a temperature
\be T_U=\frac{\hbar a}{2\pi k_B} \enspace . \ee
The common feature with the black hole case is the presence of
a {\em horizon} which in particular is responsible for the
thermal nature of the radiation. In fact, (7) directly follows from
(2) upon replacing the surface gravity $\kappa$ by $a$.
Israel (1976) showed that observers whose observations are limited
by a horizon see a ``thermal vacuum state". This follows after
summing over the unobservable states behind the horizon.
It must be emphasised that near the horizon the black hole
geometry resembles the geometry of Rindler spacetime ('t~Hooft
1996), which is the spacetime appropriate for an accelerated
observer.

For a quasistationary observer near a black hole (i.e., at
a fixed radial distance $r$ from the hole), Hawking effect and
Unruh effect are intertwined through the formula
\be T_{BH}(r)=\frac{\hbar\kappa}{2\pi k_B\chi(r)} \enspace , \ee
where $\chi(r)$ is the redshift factor of the black hole geometry,
and the spherically-symmetric case was assumed.
(A position-dependent temperature is a typical feature of
gravitational systems.)
In the limit $r\to\infty$ the Hawking effect (2) is recovered
(thermal radiation at infinity), while for $r\to R_0$, the
effect is purely one of acceleration and (7) is recovered.
This ``thermal atmosphere" near the horizon plays an important role
in many discussions of black hole entropy, see below.

An interesting connection between the Unruh effect and the
Schwinger effect (pair creation of charged particles in an external
electric field) was discussed by Parentani and Massar (1997).
This analogy enabled them to associate a formal entropy with
the Unruh effect, $S_U=k_B\pi M^2/e{\cal E}\hbar$,
 where ${\cal E}$ is the constant accelerating electric field,
and $M$ is the mass of the charged particle. With $a=e{\cal E}/M$
one has $S_U\propto M^2$ and $T_U\propto M^{-1}$, i.e. a formal analogy
to the Hawking effect (although with a different interpretation,
since here $M$ refers to the quantum field, while in the
black hole case, $m$ refers to the classical black hole mass).

\vskip 2mm
\small

Can the Unruh effect (7) be observed? Bell and Leinaas (1987)
discussed the motion of electrons in storage rings. For such circular
motion, the effect is not purely thermal, since there is no horizon.
Still, this effect leads to a change in the spin polarisation of
the electron, which may be obervable. However, present measurements
of this polarisation are not precise enough to unambigiously 
uncover such an effect from the data.

A related effect (quantum radiation by moving interfaces between
different dielectrics) could be responsible for sonoluminescence
(light emission by sound-driven air bubbles in water), which 
until now remains unexplained, see Eberlein (1996). This is
undecided at the moment.

\normalsize  

\subsection{Interpretation of Entropy}

If black holes can be attributed a genuine entropy, see (3),
the question arises whether a generalised Second Law of the kind
\be \frac{\D}{\D t}\left(S_{BH}+S_M\right)\geq 0  \ee
holds, where $S_M$ denotes the entropy of ordinary matter. 
This was investigated in many special situations, and numerous
gedankenexperimente have shown that (9) in fact holds, i.e. that there exists
no perpetuum mobile of the second kind in black hole physics. 
A typical situation is one where a box containing thermal radiation
(this maximises the matter entropy) is lowered in a quasistationary
manner towards a black hole, into which the radiation is then thrown,
see Bekenstein (1980), and Israel (this volume).
 Unruh and Wald (1982) have shown that
there is a minimal change of entropy if the box is opened at the
floating point given by the Archimedean principle (weight of box
is equal to the buoyancy from the Unruh radiation), which is just enough
to save the Second Law (9). In this discussion the relation
(8) plays an important role.

Frolov and Page (1993) have given a proof for the generalised
Second Law (9) under the assumptions that one remains within
the semiclassical approximation and that a special initial state
(no correlation between modes coming out of the past horizon
and modes coming in from past null infinity) is chosen. 
The choice of a special initial state is of course a necessary
prerequisite for any derivation of a Second Law, see Zeh (1992)
and Sect.~4.

The above discussion remains fully within the context of 
phenomenological thermodynamics (similar to discussions in the
last century before the advent of the molecular hypothesis).
A most interesting question is then whether $S_{BH}$ can be derived from 
quantum statistical considerations,
\be S_{BH}\stackrel{?}{=}-k_B\mbox{Tr}(\rho\ln\rho)
                       \equiv S_{SM} \ee
with an appropriate density matrix $\rho$. This is a key issue in
the process of understanding black holes in quantum gravity.
Does black hole entropy, for example, correspond to the large
number of states which may be hidden behind the horizon?
Or does it correspond to the large number of possible initial
states? {\em Where} is the entropy located (if at all)?
These question may indicate the kind of questions that arise.

Using a flat space example (with a surface that separates two regions
and that mimics a horizon),
Bombelli et al. (1986), and Srednicki (1993) have argued that
the entropy is located near the horizon. This may also be suggested
by the presence of the thermal atmosphere there, see the discussion
after (8). In the black hole context, this was investigated by
Frolov and Novikov (1993). They showed that by counting internal
degrees of freedom one gets $S_{SM}\propto A$.
All these authors found, however, a divergent
prefactor. Although lying inside, these degrees of freedom
are located mainly in the vicinity of the horizon.
An attempt to show that (10) can be derived from the number of possible
initial configurations of the hole was made by Zurek and Thorne (1985).

A concrete realisation of the ideas of Frolov and Novikov (1993)
was done by Barvinsky, Frolov, and Zelnikov (1995).
They consider a quantum state for the black hole and make the ansatz
that this state is constructed from the no-boundary proposal
of Hartle and Hawking (1983). The wave function is defined
on three-dimensional geometries and matter fields thereon, see
Sect.~3. The three-geometry is taken to be the Einstein-Rosen
bridge $\Sigma\equiv\bbbr\times S^2$.\footnote{It
 is shown that this state
is equal to the so-called Hartle-Hawking vacuum state which is
relevant for eternal holes, see Hartle and Hawking (1976). 
This thus provides an example where both types of ``Hartle-Hawking"
agree.} The density matrix $\rho_{in}$ of the black hole is then
obtained from this pure state by tracing out all degrees of freedom
outside the horizon. For the statistical mechanical entropy
this leads to
\be S_{SM}=-k_B\mbox{Tr}\left(\rho_{in}\ln\rho_{in}\right)=
     k_B\frac{A}{360\pi l^2} \enspace , \ee
where $l$ is a cutoff parameter (proper distance to the horizon).
One recognises that one gets a divergent result for $l\to 0$.
(Taking for $l$ the Planck length $l_P\equiv\sqrt{G\hbar}$
would yield a result proportional to (3).)
It is speculated that a finite result is obtained after the
quantum gravitational ``uncertainty" of the horizon is taken
into account, see also Sect.~4.

Since 
\[ \mbox{Tr}\left(\rho_{in}\ln\rho_{in}\right)=
    \mbox{Tr}\left(\rho_{out}\ln\rho_{out}\right) \]
(see e.g. p.~297 in Giulini et al. (1996)), the result $S\propto A$
also follows in approaches where the degrees of freedom
lie {\em outside} the horizon. An example is the ``brick wall model"
of 't Hooft (1996), see also his contribution to this volume.

The above result by Barvinsky, Frolov, and Zelnikov (1995)
arises entirely from the ``one-loop level" of the wave function
(that is the level of the WKB prefactor). Usually, however, 
$S_{BH}$, Eq. (3), is recovered solely from the classical
action, which corresponds to the ``tree level" of approximation.
Since this latter type of derivation plays a crucial role in many
discussions, and will in particular be of some relevance in
Sect.~3, a brief overview will now be given.

The origin of these discussions goes back to Gibbons and Hawking
(1977) who extended the analogy between path integrals and
partition sums to gravitational systems. This analogy, on the
other hand, was introduced within ordinary statistical mechanics
by Feynman and Hibbs (1965). 

Consider the partition sum of the canonical ensemble,
\be \E^{-\beta F}\equiv Z=\mbox{Tr}\E^{-\beta\hat{H}} \enspace , \ee
where $\beta=(k_BT)^{-1}$, and $F$ is the free energy.
On the other hand, the quantum mechanical kernel of the evolution
operator reads
\be G(x,t;x',0)=\langle x\vert\E^{-\I t\hat{H}/\hbar}\vert x'
    \rangle =\int{\cal D}x(\tau)\ \E^{\I S[x(\tau)]/\hbar}
    \enspace , \ee
where also its expression in terms of path integrals is given
(the paths going through $x'$ at time $0$ and through $x$ at
time $t$). For simplicity, I have suppressed all indices which may
be attached to $x$.

The partition sum $Z$ can be evaluated in this way, if one
transforms $t\to -\I\beta\hbar$ and performs a trace:
\be Z=\int \D x\ G(x,-\I\beta\hbar;x,0)
     = \int{\cal D}x(\tau)\ \E^{-I[x(\tau)]/\hbar} \enspace . \ee
The paths go now from $x$ at ``time" $0$ back to $x$ at ``time"
$\beta\hbar$. ($I$ denotes the euclidean action.)
 To express $Z$ in this way is especially suited
for perturbation theory, see Feynman and Hibbs (1965).
If the Hamiltonian has the standard form
\be \hat{H}=\frac{\hat{p}^2}{2m}+ V(\hat{x}) \enspace , \ee
one finds in perturbation theory (the ``small" parameter
being $\beta\hbar$)
 for the free energy
the expression (see standard books on statistical mechanics)
\be F=F_0 + \frac{\hbar^2\beta^2}{24m}\langle V'(x)^2\rangle
    \enspace , \ee
where the expectation value is performed with respect to the
canonical ensemble. The first term, $F_0$, gives the classical
value for the free energy (``tree level"). It follows from
evaluating the classical action upon classical trajectories.
Because the action contains an integration from $0$ to $\beta\hbar$,
for small $\beta\hbar$ (corresponding to $\hbar\to0$
or $T\to\infty$) the result for $F_0$ is linear in $\beta$
and independent of $\hbar$. The second term in (16) 
describes the ``one-loop level" of the perturbation. It follows from
an evaluation of the quadratic fluctuations around the classical
action. (There is no term linear in $\hbar$.)

If $Z$ (or $F$) is known, all other thermodynamic quantities
(in particular the entropy) can be calculated. The mean value
of the Hamiltonian is
\be \langle\hat{H}\rangle \equiv E=-\frac{\partial\ln Z}{\partial\beta}
     \enspace , \ee
the entropy is given by
\be S=k_B(\ln Z+\beta E)=\frac{E-F}{T}= -\frac{\partial F}
     {\partial T} \enspace . \ee
One also has $S\approx k_B\ln g(E)$, where $g(E)$ is the number
of states in the energy interval given by the mean square deviation
of the energy. The specific heats can be inferred from
second derivates of the partition sum,
\be  C= k_B\beta^2\frac{\partial^2\ln Z}{\partial\beta^2}
    =k_B(\Delta\hat{H})^2\beta^2=-\beta\frac{\partial S}{\partial\beta}
    \enspace . \ee

Gibbons and Hawking (1977) now used a (formal) quantum gravitational
path integral to evaluate the partition sum in the gravitational
context, see also Hawking (1979) and Hawking and Penrose (1996).
In contrast to the above standard context, the euclidean 
viewpoint is there assumed to be fundamental and not just a convenient
rewriting of the original lorentzian theory.      

The path integral cannot, of course, be evaluated exactly
(and it is unclear, whether it can be rigorously defined in
quantum gravity). One can, however, resort to a steepest descent
(saddle point) approximation, where only the first (and sometimes the second)
contribution is taken into account. The first contribution
is just the classical action evaluated for a classical
solution of Einstein's equations. The next order takes into
account the standard WKB-prefactor.

 The euclidean action of vacuum general relativity without
  cosmological constant reads
\be I=-\frac{1}{16\pi G}\int\D^4x\ R\sqrt{g}
    +\frac{1}{8\pi G}\int\D^3x\ (K-K^0)\sqrt{h} \enspace . \ee
In the volume term, $R$ denotes the four-dimensional Ricci scalar, and $g$
the determinant of the four-dimensional metric. In the boundary term,
$K$ denotes the trace of the extrinsic curvature, and $h$ the determinant
of the three-dimensional metric.
 For purposes of regularisation
in the asymptotically flat case,
the trace of the extrinsic curvature $K^0$ of the same boundary
embedded in {\em flat} space has to be subtracted. 

If one considers spherically symmetric uncharged black holes,
one has to evaluate (20) for the euclidean Schwarzschild solution
(the generalisation to $q\neq 0$ is straightforward).
For this solution $R=0$, and there is thus no contribution
from the volume term. The whole contribution
(which I shall call $I^*$) thus arises
from the boundary which here is the $t$-axis times a sphere of 
large radius. This is a typical feature of black hole physics,
which we shall encounter again in the course of this lecture. 

To evaluate the partition sum one has to start from the expression
(14), where one has to sum over all four-dimensional metrics
instead of just paths $x(\tau)$. In the saddle point approximation
one has (denoting with $\mbox{g}$ symbolically the four-dimensional
metric),
\be Z=\int{\cal D}\mbox{g}(x)\E^{-I[\mbox{g}(x)]/\hbar}
    \approx\exp(-I^*/\hbar)= \exp\left(-\frac{(\beta\hbar)^2}{16\pi G\hbar}
    \right) \enspace . \ee
It is due to the fact that only the {\em boundary term} of the euclidean
action contributes to (21), that the lowest order approximation
of the path integral (the ``tree level") depends already
quadratically on $\beta$. As one recognises from (16) and the
discussion following it,
in the standard situation $\beta$ occurs quadratically only at the
next order. 

{}From (17) one immediately finds
\be \langle\hat{H}\rangle =E=\frac{\hbar\beta}{8\pi G}=m
    \ee
which leads to the expression (4) for the temperature (with $q=0$).
{}From (18) one finds for the entropy
\be S=k_B(\ln Z+\beta m)= \frac{\hbar\beta^2}{16\pi G}=
    \frac{k_BA}{4G\hbar}=S_{BH} \enspace . \ee
If $Z$ had only a linear dependence on $\beta$, the entropy
would turn out to be zero.

{}From (19) one finds $C=-\beta^2\hbar/8\pi G$ and thus a {\em negative
specific heat}! This is in particular in conflict with the
positivity of $(\Delta\hat{H})^2$ und means, of course,
 that the black hole is unstable in asymptotically flat space,
as can immediately be inferred from the inverse mass dependence
of the Hawking temperature (4). As such, this is not very surprising,
since instability is typical for gravitational phenomena
(Zeh~1992). This negativity is therefore not an artifact of the
tree-level approximation.

\vskip 2mm
\small

Davies (1977) showed that for rotating or charged holes,
the specific heat can become positive for $J/m\ga 0.68Gm$
(rotating holes, where $J$ is the angular momentum) and
$q\ga 0.86Gm$ (charged holes).

\vskip 2mm
\normalsize

In the attempt to find a thermodynamically stable situation,
Gibbons and Perry (1978) considered a microcanonical ensemble
of a black hole immersed in a bath of radiation with fixed volume:
 They found that at a sufficiently
high energy density a black hole will nucleate from
a box containing radiation, in the same way as a liquid drop
can condense out of saturated vapour. However, to obtain stability
the black hole mass $m$ must be about $98\%$ of the total energy,
which means that the radiation cannot serve as a heat bath for
the hole.

In a canonical ensemble description, the specific heat can be made
positive if the black hole is put into a box (York~1986, 1991).
At the boundary of the box, boundary conditions must be specified,
i.e. in the Schwarzschild case
 one can fix the temperature of the box and its radius $r_B$. 
It follows then that stability can be achieved 
for $2Gm < r_B < 3Gm$, i.e. only for a very small box.

Alternatively, one can use a microcanonical description, where the
energy (and other extensive variables) are fixed at the boundary
(Brown and York~1993). This is very natural for gravitating
systems where energy can be expressed as a surface integral.
Instead of the euclidean path integral (14) for the canonical
partition sum, one can express the density of states $\nu(E)$
directly as a {\em lorentzian} path integral,
\[ \nu(E)= \int{\cal D}x(t)\ \E^{\I S_E[x(t)]/\hbar}
    \enspace , \]
where $S_E$ is Jacobi's action in which the energy is fixed. The sum
goes over all paths that are periodic in real time. This path integral
may be defined even in cases where the canonical partition function
(which follows via an integral transform) is divergent.
Brown and York (1993) showed that $\ln\nu\approx A/4G\hbar$, as long
as the black hole can be described semiclassically by any real
stationary axisymmetric black hole.

If the hole is charged, one must in addition fix the charge at the
boundary or, alternatively, the electric potential, see
Braden et al. (1990).

\vskip 2mm
\small

Iyer and Wald (1995) gave a comparison between
the Noether charge approach, see (1), and various euclidean
approaches. They showed that the results agree in their respective
domains of applicability, see also Brown (1995).
It is interesting that $\exp(S_{BH})$ also gives the
enhancement factor for the rate of black hole pair creation
relative to ordinary pair creation, in accordance with
the heuristic interpretation of this factor as the number of internal
states of the hole.

\vskip 2mm
\normalsize
Can these derivations of black hole entropy at the {\em tree} level
be reconciled with the above-mentioned derivations at the
{\em one-loop} level, see (11)? Problems arise due to the 
UV-divergencies connected with one-loop calculations:
For renormalisation one needs to subtract the infinite quantity
$S_{BH}(G_{bare})$ evaluated at the ``bare" gravitational constant
$G_{bare}$, a quantity that has no clear statistical mechanical meaning.
As Frolov, Fursaev, and Zelnikov (1997) have shown, this difficulty
can be avoided in theories where $G_{bare}^{-1}=0$, such as
Sakharov's induced gravity, see also Frolov and Fursaev (1998)
for a review: If one includes there non-minimally coupled scalar fields
or additional vector fields, one obtains a finite entropy
that is equal to $S_{BH}$. In induced gravity, the dynamical degrees
of freedom of the gravitational field arise from collective
quantum excitations of heavy matter fields. The same fields
produce $S_{BH}$, since the gravitational action is already itself
a ``one-loop effect". This result may also indicate why
superstring theory, another ``effective theory of gravity",
allows one to reproduce $S_{BH}$ from the counting of
quantum states, see Sect.~4.

It was the intention of this section to give convincing arguments
that black holes must be quantum objects and that they can be
fundamentally understood only in the framework of quantum gravity.
Before I shall discuss some approaches to quantum gravity
in more detail, I want to remark that one can already speculate
from the above results about some possible features of the
full theory. One result of such a speculation is the intriguing
feature of a possible area (and thus mass) quantisation for
a black hole, see e.g. Bekenstein (1997), and the references 
therein. It was suggested from heuristic considerations that
\be A=16\pi(Gm)^2=4G(\ln 2)\hbar n, \quad n\in\bbbn \enspace .\ee
This would already in the semiclassical theory change drastically
the spectrum of black hole radiation. For example, no quanta
would be emitted with frequencies lower than some fundamental
frequency $(\ln 2)/8\pi Gm$, in contrast to the thermal nature
of Hawking radiation. One could thus test this effect of quantum gravity
already for $m\gg m_P$ (provided that primordial holes exist).

The result (2) of a {\em thermal} spectrum of black hole
radiation was obtained in the semiclassical limit, where
gravity is treated classically. If it were true even in the full
theory of quantum gravity, it would mean that ``information"
were lost in the following sense: Since one can in principle
start from any initial quantum state (even a pure one),
its exact evolution into a thermal state would contradict
the unitary evolution law of standard quantum theory.
In this case, a theory of quantum gravity would possess some
radical new features. Since, however, the full theory is not
yet known, the answer to this {\em problem of information
loss} is also not yet known (see, for example, the
review in Giddings~1994). This ``problem" may, however, serve as
a useful leitmotif in the search for a full theory. How even the
semiclassical limit might be altered has been mentioned in
the context of (24). The effect of quantum gravitational
corrections on this information loss will be briefly discussed
in Sect.~3.2.

\section{Black Holes in Canonical Quantum Gravity}
\subsection{A Brief Introduction into Canonical Gravity}

Canonical quantum gravity is obtained via the application
of standard canonical quantisation rules to the theory of
general relativity (or some other classical theory, but I shall restrict
myself to general relativity). Since this does not provide a unified 
description of all fields, it is expected that the resulting
framework is only an effective theory. There is, however, the hope
that canonical quantum gravity may reflect many of the features
of a genuine quantum theory of gravity. Its formulation must be
intrinsically non-perturbative, since general relativity is known
to lead to a non-renormalisable quantum theory at the perturbative
level.
A perhaps more serious candidate for a genuine
quantum theory of gravity
unifying all interactions, superstring theory, is
briefly described in the next section.

The canonical framework assumes that the classical spacetime ${\cal M}$
is globally hyperbolic, ${\cal M}=\Sigma\times\bbbr$, such that 
a $3+1$ decomposition (a foliation into spacelike hypersurfaces)
can be performed. This is already of relevance for the
classical theory because it allows one to pose a well-defined
Cauchy problem (e.g. in numerical relativity, see the
contribution of Seidel to this volume). A $3+1$ formulation
is required because the canonical approach is a Hamiltonian
formulation of the theory.
 Due to the
four-dimensional diffeomorphism invariance (``coordinate
invariance" in spacetime), the classical
theory contains four constraints at each space point,
one Hamiltonian constraint,
\be  {\cal H} \approx 0, \enspace \ee
and three spatial diffeomorphism constraints (``coordinate invariance"
on the three-dimensional spatial hypersurface $\Sigma$),
\be {\cal D}_a \approx 0. \enspace \ee
Here, as usual, $\approx$ denotes the weak equality in the
sense of Dirac.

The canonical configuration variable can
be chosen to be the three-dimensional metric
$h_{ab}(\vec{x})$ on $\Sigma$, and the canonical momentum is then a linear
function of the extrinsic curvature of $\Sigma$. To this one can add
any matter fields in the standard manner. This constitutes the
traditional, geometrodynamic, approach. Alternatively, one may choose
a complex connection or so-called 
loop variables on $\Sigma$ for the configuration
variables. This brings in many formal similarities to
Yang--Mills theories. I want to emphasise that the constraint structure
(25, 26) is typical for all versions of canonical theories that possess a
diffeomorphism invariance on the classical level, even
if the specific form is different. This is the basis for the hope
that these versions have important common features.
Also superstring theory has a constraint structure, although
its interpretation is somewhat different from here.

In the following I want to restrict myself to the quantisation
method proposed by Dirac. This means to {\em formally} transform
the above constraint equations into operator equations acting
on physical states $\Psi$, 
\be \hat{\cal H}\Psi =0 \enspace , \ee
and
\be \hat{\cal D}_a\Psi =0 \enspace . \ee
The wave functional $\Psi$ depends, in the geometrodynamic approach,
on the three-metric (as well as on non-gravitational fields),
in the other approaches mentioned above on the complex connection
or on loop variables.\footnote{In the latter cases there are also
additional constraints coming from triad rotations.}
Due to the constraints (28), the wave functional is invariant under
three-dimensional coordinate transformations. This is often
indicated by writing $\Psi[^3{\cal G}]$, where $^3{\cal G}$ means
``three-geometry", although this is a loose notation, since $\Psi$
cannot explicitly be given in this form.
 
If space is compact, there are no further
constraints. If not, additional constraints arise from variables
living at boundaries. This will be of particular relevance for
our treatment of black holes, see Sect. 3.2.

It cannot be the purpose of this article to give a detailed introduction
into this approach and its problems. A comprehensive reference is
Ehlers and Friedrich (1994). A recent report on the connection
and loops approaches can be found, for example, in Ashtekar (1997);
a recent report on conceptual problems in Isham (1997). 
A comprehensive review of canonical quantum gravity as applied to
cosmology is Halliwell (1991). The black hole examples discussed
below may also be thought to give illustrative examples
for the full framework.
 
A helpful analogy between ordinary (quantum) mechanics and
(quantum) general relativity is given in Table~2.

\begin{table}[htb]
\caption[ ]{Comparison of mechanics and general relativity}
\begin{flushleft}
\renewcommand{\arraystretch}{1.2}  
\begin{tabular}{l|l}
{Mechanics of one particle} &\ {General relativity} \\ \hline
{} & {} \\
position $q$ &\  geometry $^3{\cal G}$ of a  \\
{}  & \ three-dimensional space\\
{} & {} \\
trajectory $q(t)$ &\ spacetime $\{^3{\cal G}(t)\}\equiv\ ^4{\cal G}$ \\
{} & {} \\
uncertainty between &\ uncertainty between\\
position and momentum &\ ``space and time"\\
{} & \ (three-geometry and extrinsic curvature)\\
{} & {} \\
$\psi(q,t)$ &\ $\Psi[^3{\cal G},t)\equiv \Psi[^3{\cal G}]$ \\
\end{tabular}
\renewcommand{\arraystretch}{1}
\end{flushleft}\end{table}

The most important conceptual lesson from the above comparison
is that spacetime has no fundamental meaning in canonical quantum
gravity, in the same way as a particle trajectory has no 
fundamental meaning in quantum mechanics.
This fact lies behind the so-called ``problem of time" in quantum
gravity -- the absence of any external time parameter in the constraint
equations (27, 28), and the related problem of which Hilbert space
(if any) to choose in quantum gravity. (This is way the
quantum gravitational wave function in Table~2 is $t$-independent.)
 To a large extent, these issues
are open, see e.g. Kiefer (1997b). Fortunately, in the
black hole case, the ``rest of the Universe" can be assumed
to be in a semiclassical regime where a concept of time
exists, so that some of the above conceptual problems don't have
to be dealt with in the first place. These problems are, however,
relevant if the whole Universe including the black hole is described
in quantum terms, see Sect.~4.

A frequently employed approximation scheme is to perform a semiclassical
expansion of the equations (27, 28), see Kiefer (1994).
One writes the full wave functional as $\Psi\equiv\exp(\I S/\hbar)$
with some arbitrary complex function $S$ which is expanded into
powers of the gravitational constant: $S=G^{-1}S_0+S_1
+GS_2+\ldots$. This is then inserted into (27, 28), leading to
equations at consecutive orders of $G$. It must be emphasised that
this can be done only in a formal way, since it is unclear how 
to rigorously define the equations (27, 28). For finite-dimensional
models it was shown by Barvinsky and Krykhtin (1993) and Barvinsky
(1993) how up to ``one loop" a consistent factor ordering
and a consistent Hilbert space structure can be obtained.
The important open issue is to find a consistent, anomaly-free,
regularisation for their equations in the field theoretic case.

 The highest order ($G^1$)
yields the gravitational 
Hamilton-Jacobi equation for $S_0$. This is equivalent 
to the classical Einstein equations and corresponds to the
``tree level" of the theory. A special solution $S_0$ thus corresponds
to a family of classical spacetimes. 
The next order ($G^0$) leads to a functional Schr\"odinger equation
for non-gravitational fields in a given background. 
It corresponds to the ``one-loop"
 limit of quantum field theory in an external
background, the limit in which the Hawking radiation is derived.
Higher orders in $G$ then lead to genuine quantum gravitational
correction terms as well as back reaction terms from the
non-gravitational fields onto the semiclassical background.

\vskip 2mm
\small

The approximation scheme sketched above is not unique.
Alternative schemes can be found, e.g., in Bertoni, Finelli, and Venturi
(1996), and Kim (1997). They differ from the above in the treatment
of the back reaction of the non-gravitational fields.

\vskip 2mm
\normalsize

The next section is devoted to the application of canonical methods
to a particular situation: spherically symmetric black holes.

\subsection{Quantisation of spherically symmetric black holes}

The first model which I shall briefly describe is the 
case of spherically symmetric black holes. I shall begin with
the so-called ``eternal hole", where only the gravitational
degrees of freedom (and, in the Reissner-Nordstr\"om case,
the electromagnetic field) are taken into account. The more 
realistic case where additional dynamical fields (such as
a scalar field) are present is discussed thereafter.
 
The eternal Schwarzschild hole was first discussed by Thiemann
and Kastrup (1993), Kastrup and Thiemann (1994) within the
connection dynamical approach and then by Kucha\v{r} (1994)
in the geometrodynamical approach, see also
Cavagli\`{a}, de Alfaro, and Filippov (1996).
I shall follow the
geometrodynamical approach and generalise it to include the
Reissner-Nordstr\"om case, see also Louko and Winters-Hilt (1996).
``Eternal" refers to the time-symmetric case where both a past and a future
horizon are present (``complete Kruskal diagramme").
Such holes cannot result from a collapse. Although thus being
unrealistic from an astrophysical point of view, eternal holes
provide a useful (and relatively simple) framework for questions
of principle. 

Starting point is the ADM form for general spherical symmetric
metrics on $\bbbr\times\bbbr\times S^2$:
\be \D s^2= -N^2(r,t)\D t^2 + \Lambda^2(r,t)(\D r+N^r(r,t)\D t)^2+
    R^2(r,t)\D\Omega^2 \enspace . \ee
The lapse function $N$ encodes the possibility to perform arbitrary
reparametrisations of the time parameter, while the shift function
$N^r$ is responsible for reparametrisations of the radial
coordinate (this is the only freedom in performing spatial
coordinate transformations that is left after spherical symmetry has been
imposed). The parameter $r$ is only a label for the spatial
hypersurfaces; if the hypersurface extends from the left to the
right wedge in the Kruskal diagramme, one takes $r\in(-\infty,
\infty)$. If the hypersurface originates at the bifurcation point
where path and future horizon meet, $r\in(0, \infty)$. 
If one has in addition a spherically symmetric electromagnetic field,
one makes the following ansatz for the one-form potential:
\be A=\phi(r,t)\D t+\Gamma(r,t)\D r \enspace . \ee
In the Hamiltonian formulation, $\phi$ as well as $N$ and $N^r$
are Lagrange multipliers whose variations yield the constraints of the
theory. Variation of the Einstein-Hilbert action
 with respect to $N$ yields the Hamiltonian
constraint (25) which for the spherically symmetric model reads
\be {\cal H}= \frac{G}{2}\frac{\Lambda P_{\Lambda}^2}{R^2}
  -G\frac{P_{\Lambda}P_R}{R}+ \frac{\Lambda P_{\Gamma}^2}
    {2R^2}+G^{-1}V_G \approx 0 \enspace , \ee
where the gravitational potential term reads, explicitly,
\be V_G= \frac{RR''}{\Lambda}- \frac{RR'\Lambda'}{\Lambda^2}
   +\frac{R'^2}{2\Lambda}-\frac{\Lambda}{2} \enspace . \ee
(A prime denotes differentiation with respect to $r$.)
Variation with respect to $N^r$ yields one (radial) diffeomorphism
constraint (26),
\be {\cal D}_r= P_RR'- \Lambda P_{\Lambda}'\approx 0 \enspace . \ee
One recognises from this constraint that $R$ transforms as
a scalar, while $\Lambda$ transforms as a scalar density.

Variation of the action with respect to $\phi$ yields as usual the Gau\ss\
constraint
\be {\cal G}= P_{\Gamma}'\approx 0 \enspace . \ee
The constraint (33) generates radial diffeomorphisms for the fields
$R$, $\Lambda$ and their canonical momenta. It does not
generate diffeomorphisms for the electromagnetic variables.
This can be taken into account if one uses the multiplier
$\tilde{\phi}=\phi-N^r\Gamma$ instead of $\phi$ and varies
with respect to $\tilde{\phi}$
(Louko and Winters-Hilt~1996), but for our purposes it is sufficient
to stick to the above form (33).

The model of spherical symmetric gravity can be embedded into a
whole class of models usually referred to as ``two-dimensional
dilaton gravity theories". This terminology comes from effective
two-dimensional theories (usually motivated by string theory)
 which contain in the gravitational sector
 a scalar field (the ``dilaton") in addition to
the two-dimensional metric (of which only the conformal factor
is relevant). 
Interest in such models arose after Callan et al. (1992) studied
one model in detail (now called the CGHS model), in which they
addressed the issues of Hawking radiation
 and back reaction\footnote{A detailed review
 of two-dimensional black holes
is Strominger (1995).}.
This was facilitated by the fact that this model is classically
soluble even if another, conformally coupled, scalar field
is included. The canonical formulation of this model can be found,
e.g., in Louis-Martinez, Gegenberg, and Kunstatter (1994) and
Demers and Kiefer (1996). The dilaton field is analogous
to the field $R$ from above, while the conformal factor of the
two-dimensional metric is analogous to $\Lambda$.

\vskip 2mm
\small
The dilaton model contains one non-trivial parameter, the
constant $\lambda$ which has the dimension of an inverse length.
The corresponding Hawking temperature and entropy are given by,
respectively,
\[ T_{BH}=\frac{\hbar\lambda}{2\pi k_B},\quad
   S_{BH}=\frac{2\pi k_B m}{\hbar\lambda}\ . \]
Note that the temperature is here independent of the black hole
mass $m$, and that therefore the entropy is linear in $m$.
This is also the reason why some aspects of this models
are unrealistic from the four-dimensional point of view.

\vskip 2mm
\normalsize

Coming back to the spherically symmetric model, consider first
the boundary conditions for $r\to\infty$. (If $r\in(-\infty,\infty)$,
there are analogous conditions for $r\to-\infty$ which will be
ignored here, see Kucha\v{r} (1994).) For $r\to\infty$ one has
in particular
\be \Lambda(r,t)\to 1+\frac{Gm(t)}{r},\; R(r,t)\to r,\;
    N\to N(t) \enspace, \ee
as well as
\be P_{\Gamma}(r,t)\to q(t), \quad \phi(r,t)\to \phi(t) \enspace .\ee
{}From the variation with respect to $\Lambda$ one then finds the
boundary term \\ $G\int \D t\ N\delta m$. In order to avoid the unwanted
conclusion $N=0$ (no evolution at infinity), one has to
compensate this term in advance by adding the boundary term
\[ -G\int\D t\ Nm \]
to the classical action. Note that $m$ is just the ADM mass.
The need to include such a boundary term was recognised by
Regge and Teitelboim (1974).
Similarly, for charged holes, one has to add the term
\[ -\int\D t\ \phi q \]
to compensate for $\int \D t\ \phi\delta q$ which arises from
varying $P_{\Gamma}$. If one wished instead to consider $q$ as a given,
external parameter, this boundary term would be obsolete.

As long as restriction is made to the eternal hole,
appropriate canonical transformations allow to simplify the classical
constraint equations considerably (Kucha\v{r} 1994,
Louko and Winters-Hilt~1996). One gets
\[ \left(\Lambda, P_{\Lambda}; R, P_R;
    \Gamma, P_{\Gamma}\right) \longrightarrow
    \left({\cal M}, P_{{\cal M}}; {\cal R}, P_{{\cal R}};
   Q, P_Q\right) \enspace . \]
In particular,
\bea {\cal M}(r,t)&=& \frac{P_{\Gamma}^2+P_{\Lambda}^2}{2R}
   +\frac{R}{2}\left(1-\frac{R'^2}{\Lambda^2}\right)
   \stackrel{r\to\infty}{\longrightarrow} m(t)  \\
    Q(r,t) &=& P_{\Gamma} \stackrel{r\to\infty}
    {\longrightarrow} q(t) \enspace . \eea
(I note that ${\cal R}=R$ and that the expression for
$P_{\cal R}$ is somewhat lengthy and will not be given here.)

The new constraints, which are equivalent to the old ones, read
\bea  {\cal M}'=0 \quad &\Rightarrow& \quad {\cal M}(r,t)=m(t), \\
      Q'=0 \quad &\Rightarrow& \quad Q(r,t)= q(t),  \\
      P_{{\cal R}}=0 & &\enspace . \eea  
Note that $N(t)$ and $\phi(t)$ are prescribed functions that must
not be varied; otherwise one would be led to the unwanted
restriction that $m=0=q$. This can be remedied if the action is
parametrised, bringing in new dynamical variables,
\bea  N(t) &=:& \dot{\tau}(t), \nonumber\\
      \phi(t) &=:& \dot{\lambda}(t) \enspace . \eea
Here, $\tau$ is the proper time that is measured with standard
clocks at infinity, and $\lambda$ is the variable conjugate
to charge; $\lambda$ is therefore connected with the elctromagnetic
gauge parameter at the boundaries.
 In the canonical formalism one has to introduce momenta
conjugate to these variables, which will be denoted
$\pi_{\tau}$ and $\pi_{\lambda}$, respectively.
This, in turn, requires the introduction of additional constraints
linear in momenta,
\bea {\cal C}_{\tau} &=& \pi_{\tau} +Gm\approx 0,  \\
     {\cal C}_{\lambda} &=& \pi_{\lambda} +q \approx 0 \eea
which have to be added to the action:
\bea -G\int\D t\ m\dot{\tau} \quad &\to& \quad
     \int\D t\ (\pi_{\tau}\dot{\tau}-N{\cal C}_{\tau}), \\
    -\int\D t\ q\dot{\lambda} \quad &\to& \quad
     \int\D t\ (\pi_{\lambda}\dot{\lambda}-\phi{\cal C}_{\lambda})
     \enspace . \eea
The remaining constraints in this model are thus (41) and (43,44).

Quantisation proceeds then in the way sketched in Sect.~3.1 by
acting with an operator version of the constraints on wave functionals
$\Psi[{\cal R}(r);\tau,\lambda)$. Since (41) leads to
$\delta\Psi/\delta{\cal R}=0$, one is left with a purely
quantum {\em mechanical} wave function $\psi(\tau,\lambda)$.
The implementation of the constraints (43,44) then yields
\bea \frac{\hbar}{\I}\frac{\partial\psi}{\partial\tau}
      +m\psi &=& 0, \\
     \frac{\hbar}{\I}\frac{\partial\psi}{\partial\lambda}
     +q\psi &=& 0 \eea
which can readily be solved to give
\be \psi(\tau,\lambda)= \chi(m,q)\E^{-\I(m\tau+q\lambda)/\hbar} \ee
with an arbitrary function $\chi(m,q)$. 
Note that $m$ and $q$ are here considered as being fixed.
The reason for this is that up to now we have restricted attention
to one semiclassical component of the wave function only
(eigenstates of mass and charge).
Superpositions of states with different $m$ and $q$ can be made,
and I shall make some remarks on this below.

If the hypersurface goes through the whole Kruskal diagramme
of the eternal hole, only the boundary term at $r\to\infty$
(and an analogous one for $r\to-\infty$) contributes. 
Of particular interest in the black hole case, however, is the case
where the surface originates at the {\em bifurcation surface}
($r\to 0$) of past and future horizons. This makes sense since
data on such a surface suffice to construct the whole right
Kruskal wedge, which is all that is accessible to an observer in this
region. Moreover, this mimics the situation where a black hole
is formed by collapse, in which the regions $III$ and $IV$ of the
Kruskal diagramme are absent.

What are the boundary conditions that are adopted at
$r\to 0$? They are chosen in such a way that the classical
solutions have a nondegenerate horizon and that the hypersurfaces
begin at $r=0$ asymptotic to hypersurfaces of constant
Killing time (Louko and Whiting~1995).
In particular, 
\bea N(r,t) &=& N_1(t)r+{\cal O}(r^3), \\
     \Lambda(r,t) &=& \Lambda_0(t)+ {\cal O}(r^2), \\
     R(r,t) &=& R_0(t)+R_2(t)r^2 +{\cal O}(r^4) \enspace . \eea
Variation leads, similarly to 
the situation at $r\to\infty$, to a boundary term at $r=0$:
\[ -N_1R_0(G\Lambda_0)^{-1}\delta R_0 \enspace . \]
If $N_1\neq0$, this term must be subtracted 
($N_1=0$ corresponds to the case of extremal holes,
$\vert q\vert=Gm$, which is characterised by 
$\partial N/\partial r(r=0)=0$.)
Introducing the notation $N_0\equiv N_1/\Lambda_0$, the boundary term
to be added to the classical action reads
\[  (2G)^{-1}\int\D t\ N_0R_0^2 \enspace . \]
The quantity
\be \alpha\equiv \int_{t_1}^{t}\D t\ N_0(t) \ee
can be interpreted as a ``rapidity" because it boosts the normal vector
to the hypersurfaces $t=constant$, $n^{a}$, in the way described by
\be n^{a}(t_1)n_a(t)=-\cosh\alpha \enspace , \ee
see Hayward (1993). To avoid fixing $N_0$, one introduces
an additional parametrisation (Brotz and Kiefer~1997)
\be N_0(t)=\dot{\alpha}(t) \enspace . \ee
Similarly to (45,46) above, one must replace in the action
\be (2G)^{-1}\int\D t\ R_0^2\dot{\alpha} \quad \to \quad
     \int\D t\ (\pi_{\alpha}\dot{\alpha}-N_0{\cal C}_{\alpha})
     \enspace , \ee
with the new constraint
\be {\cal C}=\pi_{\alpha}-\frac{A}{8\pi G} \approx 0 \enspace , \ee
where $A=4\pi R_0^2$ is the surface of the bifurcation sphere.
One notes that $\alpha$ and $A$ are canonically conjugate
variables, see Carlip and Teitelboim (1995).

Quantisation then leads to (taking all constraints into account)
\be \psi(\alpha,\tau,\lambda)= \chi(A,m,q)
    \exp\left(\frac{\I}{\hbar}\left[\frac{A}{8\pi G}\alpha
     -m\tau-q\lambda\right]\right) \enspace . \ee
Since $A$ occurs in the state (58), one may suspect that also
the entropy comes into play here, see (3). However, (58) is a pure
quantum state, which possesses vanishing entropy, and
$A$ is only part of its phase. The relation to entropy can only be
achieved after an appropriate euclideanisation is performed, compare
Sect.~2.3. This will be done below.
(The wave function for a Reissner-Nordstr\"om hole, if an
additional complex scalar field is coupled, can be found
in Moniz (1997). In contrast to our model, his situation
describes a dynamical evolution.)

The classical equations are found from (58) in the standard way by
finding the extremum of the phase with respect to the parameters.
For this to work, only two of the three parameters $A,m,q$
can be considered as independent. (I shall choose $m$ and $q$.)
Differentiating the phase with respect to $m$ and setting the result to zero
yields
\be \alpha= 8\pi G\left(\frac{\partial A}{\partial m}\right)^{-1}
    \tau \enspace . \ee
{}From Table~1 one recognises the occurrence of the surface gravity
$\kappa$ on the right-hand side of (59):
\be \alpha=\kappa\tau \enspace , \ee
which is just the classical relation for the rapidity, see Brotz (1997).
This is not surprising since it is known that boundary terms
in the classical action are important in the derivation of
the First Law of black hole mechanics (Wald~1997a).
Generally, conjugate quantities in thermodynamics (extensive --
intensive) correspond to conjugate variables in the Hamiltonian
formalism.

Differentiating the phase of (58) with respect to $q$
and setting the result to zero
yields
\be \phi=\frac{\kappa}{8\pi G}\frac{\partial A}{\partial q}
        =-\frac{\partial m}{\partial q}\vert_A =-\frac{q}{R_0}
    \enspace , \ee
another ``thermodynamical" relation.\footnote{It corresponds to
$\partial S/\partial N=-\mu/T$ with $\mu=-\phi$ and
$N=q$ ($N$ is the particle number and $\mu$ the chemical potential.)}
This completes the solution of the eternal Reissner-Nordstr\"om
hole. 

I shall now turn to the more realistic case where an additional
dynamical field is present. This can be used to ``form" the black hole
in the first place, and leads to the emergence of 
interesting features such as Hawking radiation. It also provides 
an interesting application of the semiclassical expansion
presented in Sect.~3.1. I denote the scalar field  by $f$, see e.g.
Romano (1995), Demers and Kiefer (1996)
 and Kucha\v{r} et al. (1997) for details of the formalism.

At order $G^0$, the total wave functional is of the form
\be \Psi\approx C^g\E^{\I S_0^g/\hbar}\bar{\chi}\enspace , \ee
where $C^g$ and $S_0^g$ depend only on the gravitational
(and electromagnetic) variables. These variables comprise
the functions $\Gamma(r), R(r), \Lambda(r)$ as well as the
boundary variables $\alpha,\tau,\lambda$. The functional $\bar{\chi}$
depends, in addition, on the scalar field $f$. The important point
is that $\bar{\chi}$ obeys a functional Schr\"odinger equation
with respect to the background found from $S_0^g$.

As in the general case, $S_0^g$ obeys the Hamilton-Jacobi equation
for gravity (plus electromagnetic field). An explicit solution
reads (Brotz and Kiefer~1997)
\bea S_0^g &=& \int_0^{\infty}\D r\left(q\Gamma +G^{-1}\Lambda F
     -G^{-1}\frac{RR'}{2}\ln\frac{R'/\Lambda +F/R}
       {R'/\Lambda -F/R}\right) \nonumber\\
    & & \; +\frac{A\alpha}{8\pi G} -m\tau -q\lambda \enspace , \eea
where
\be F=R\sqrt{\frac{R'^2}{\Lambda^2}+ \frac{2M(r)}{R} -1} \ee
and
\be M(r)=m-\frac{q^2}{2R(r)} \enspace . \ee
Note that $S_0^g$ depends parametrically on $m$ and $q$
which are just the mass and the charge of the hole, respectively.
Expression (65) is nothing but the total energy of the hole.
Inspection of (63) exhibits that the electromagnetic part in
(62) from $S_0^g$ is given by
\[ \exp\I\left(\int_0^{\infty}\D r\Gamma -\lambda\right) \enspace . \]
This expression can be understood as follows. The electromagnetic
potential (30) changes under a gauge transformations according to
\be A \to \phi \D t +\Gamma\D r +\D \xi
     =(\phi+\dot{\xi})\D t+(\Gamma+\xi')\D r \enspace . \ee
Therefore,
\[ \int_0^{\infty}\D r\Gamma(r) \to \int_0^{\infty}
    \D r\Gamma(r)+\xi(\infty)-\xi(0) \enspace . \]
Now, $\xi(\infty)-\xi(0)$ may be absorbed into $\lambda$, since
$\lambda$ itself was interpreted as the boundary gauge parameter.

Since the full theory is linear, one can perform arbitrary
superpositions of states (62) with {\em different} values
for $m$ and $q$. These describe situations where the hole has neither
a definite charge nor a definite mass. However, such superpositions
can only be distinguished from a corresponding mixture if one
could ``measure" the variables conjugate to $m$ and $q$,
i.e. $\tau$ and $\lambda$. Otherwise, effective ``superselection
rules" would result, see Giulini, Kiefer, and Zeh (1995),
and Chap.~6 of Giulini et al. (1996).

Another interesting situation is described by a superposition
of the state (62) with its complex conjugate (this is possible
since the full Wheeler-DeWitt equation is real). Such
superpositions may follow in a natural way from appropriate
boundary conditions (Hajicek~1992). It was shown in
Demers and Kiefer (1996) that these superpositions (which can be
heuristically interpreted as representing a superposition
of a black hole with a white hole) become indistinguishable
locally from a mixture after the irreversible interaction
with the Hawking radiation is taken into account -- a process
known as decoherence (Giulini et al.~1996).

How can the Hawking radiation be found from a state such as (62)?
This was clarified in Demers and Kiefer (1996) in the context
of dilaton gravity (the extension to spherically symmetric
gravity should be straightforward). One solves the functional
Schr\"odinger equation obeyed by $\bar{\chi}$ in a background
describing the collapse to a black hole. The initial state
is taken to be a Gaussian (a ``vacuum state"). During the evolution,
this state remains a Gaussian, but with a different ``width".
This just expresses the fact, as mentioned in Sect.~2.2,
that the notion of a vacuum becomes ambiguous in such a situation.
Using the initial state as the reference vacuum state also at
late times, the evolved state contains ``particles" with respect
to that vacuum. One has
\be \langle\bar{\chi}\vert\hat{n}(k)\vert\bar{\chi}\rangle
     =\frac{1}{\exp\left(\frac{\hbar\vert k\vert}{k_BT_{BH}}\right)
      -1}\enspace , \ee
where $\hat{n}$ denotes the ``particle number operator"
for the mode of wave number $k$ with respect to the
original vacuum. Note that, although $\bar{\chi}$ is a
{\em pure} quantum state, the expectation value (67) is a Planckian
distribution with respect to the Hawking temperature $T_{BH}$.
The difference of $\bar{\chi}$ to a genuine mixture will be noticed
if other expectation values (of ``higher order operators")
are performed.

For the important case where the surfaces are fixed at the
bifurcation sphere, it turns out that the field $f$ must vanish
at this point for the state $\bar{\chi}$ to be normalisable.
Thus, the bifurcation sphere acts like a ``mirror" for this field.
This is why the quantum state turns out to be a pure one.
Other surfaces which penetrate the interior of the hole
lead to a mixed state outside after the interior degrees of
freedom are ``traced out" (as in Israel~1976).

Can one go beyond the order of approximation (62)?
This is in fact possible, but so far only in a formal way,
without addressing in detail the issue of regularisation
(Kiefer~1994). Still, however, qualitative features can be studied.
At oder $G^1$, correction terms to the functional Schr\"odinger
equation obeyed by $\bar{\chi}$ are obtained. Among these terms,
there is an imaginary term, $\I\mbox{Im}H_m$, contributing
to the effective matter Hamiltonian. In the case of collapse to a
black hole, $\mbox{Im}H_m<0$ (Kiefer, M\"uller, and Singh~1994).
Since the following equation holds for the density matrix $\rho$,
\be \frac{\D}{\D t}\left([\mbox{Tr}\rho]^2-\mbox{Tr}\rho^2\right)
    =4\mbox{Tr}\left([\rho\mbox{Tr}\rho-\rho^2]\mbox{Im}H_m\right)
    \enspace , \ee
one finds from $\mbox{Im}H_m<0$ that the difference between
$(\mbox{Tr})^2$ and $\mbox{Tr}\rho^2$ decreases, corresponding
to an increase in ``purity" for the quantum state. 
Whether this may indicate a quantum gravitational ``recovery of
information" from the hole can of course only be judged from
the full, as yet elusive, theory. This result at least 
demonstrates what kind of effects one might expect to see
in higher orders of the semiclassical approximation.

At order $G^1$, also back reaction terms from the matter fields
(here from the $f$-field) onto the gravitational background
are found (Kiefer~1994). These can be evaluated only in special cases,
for example in the toy model of a 2+1-dimensional black hole
coupled to a conformal scalar field (Brotz~1998).

An interesting point is of course whether there are situations
where the semiclassical approximation breaks down in the first
place. This would mean that quantum gravity effects can become
important below the Planck scale. Keski-Vakkuri et al. (1995), for example,
arrived at the conclusion that the semiclassical approximation
breaks down at the black hole horizon, in the sense that tiny
fluctuations of the black hole mass may produce an immense
change in the matter state. The physical implications of this
result are not yet fully clear. It can also not be excluded that
anomalies in quantum gravity spoil the above semiclassical limit
and demand for an explicit modification of the constraints,
see e.g. Cangemi, Jackiw, and Zwiebach (1996).

I emphasised above that there is not yet any connection with a
notion of entropy for the pure quantum state (58). This can be
established after some ``euclideanisation" is performed,
see the discussion in Sect.~2.3. How does this work?
{}From (55) it is clear that the rapidity $\alpha$ is connected
with the lapse function. Therefore, going to the euclidean regime
means both $\tau\to-\I\beta\hbar$, see (14), and $\alpha\to
-\I\alpha_E$. Regularity of the line element then demands
that $\alpha_E=2\pi$ (Brotz and Kiefer~1997).
Consequently, the euclidean version of the quantum state (62)
contains the term
\be \exp\left(-\beta m+\frac{A}{4\hbar G}\right) \enspace . \ee 
There is in addition the euclideanised version of the integral
in (63) and the term containing $\lambda\to\lambda_E=-\I\hbar\beta\phi$.

This does of course not yet yield a partition sum.
However, after the whole semiclassical part is evaluated
at the classical value for the Hamilton-Jacobi functional
and a trace is performed, one finds by applying (17) that the
second term in (69) is just the Bekenstein-Hawking entropy (3).
Alternatively, one can interpret (69) as directly giving the
enhancement factor for the rate of black hole pair creation
relative to ordinary pair creation. Here my focus was just to show how
the expression for $S_{BH}$ emerges in the canonical
 formalism.\footnote{Due to Smarr's formula, (69) is consistent
with (21).}

Consider now the case of an extremal hole, where $\vert q\vert =Gm$.
As can be immediately inferred from the discussion after (52),
there is no surface term to consider, since $N_0=0$.
Thus, $\alpha=0$, and there is no $A$-term in (58).
This would also mean that the entropy is zero.
Recalling our discussion in Sect.~2.1, this shows that Planck's
version of the Third Law is fulfilled. This result was also
found in a variety of other approaches, see the references
in Brotz and Kiefer (1997). It is {\em not} fulfilled in string theory,
where $S_{extreme}=A/4\hbar G$, see Sect.~4.
It is also not fulfilled for the extreme (Kerr) black hole which
occurs in the transition from the disk of dust solution to
the rotating black hole solution, see Neugebauer's contribution
to this volume.

The above derivation of entropy via boundary terms suggests the
following natural interpretation in terms of ``missing
information". For surfaces which in the classical spacetime
correspond to slices through the {\em full} Kruskal diagramme, this
``information" is maximal in the sense that one can recover 
the full spacetime from data on this surface. Since no boundary
(except at infinity) is present, the entropy is zero.
For slices that start at the bifurcation sphere, this information
is less than maximal for Schwarzschild black holes and for
non-extreme Reissner-Nordstr\"om black holes. They are therefore
attributed the entropy $A/4\hbar G$. In contrast, the maximum
information (for the full spacetime up to the Cauchy horizon)
is already available for such slices in the
extreme case, as can be easily recognised from the corresponding
Penrose diagramme. Extreme holes are therefore attributed
a vanishing entropy. A somewhat related interpretation was given
in the path integral framework by Martinez (1995).
An interesting point was raised by Ghosh and Mitra (1997)
who argued that $S_{extreme}\neq0$ follows from extremisation
after quantisation, while $S_{extreme}=0$ holds for 
extremisation before quantisation.

Can the quantisation of mass (or area), as described by (24),
be found within the canonical formalism? This is, unfortunately, an 
open issue. One can, for example, postulate Bohr-Sommerfeld type
of quantisation rules in the euclidean theory (Kastrup~1996).
This would lead to
\be nh=\oint\pi_{\alpha}\D\alpha= \int_0^{2\pi}\frac{A}{8\pi G}
        \D\alpha=\frac{A}{4G}\enspace . \ee
This is similar to (24), albeit with a different numerical factor.
Whether a similar result can be found in the physically relevant
lorentzian theory remains open.

\vskip 2mm
\small
Other interesting developments can only be mentioned here.
Carlip (1997) was able to give a statistical mechanical origin
for the black hole entropy in the case of a 2+1-dimensional 
black hole. There it results from ``would-be-gauge" degrees
of freedom becoming dynamical at the horizon.
Using the loop approach to canonical quantum gravity,
Rovelli (1996) found that $S_{BH}\propto A$, although
with a numerical coefficient different from (3).

\vskip 2mm
\normalsize
To summarise, canonical quantum gravity can offer the tool
to understand quantum features of black holes such as entropy and
Hawking radiation. Still, however, the main problems are not yet
solved: Can the Bekenstein-Hawking entropy for four-dimensional
black holes be derived by counting appropriate degrees of freedom?
What is the final evolution of a black hole, after the
semiclassical approximation breaks down?


\section{Further Developments}

In Sect.~3 I discussed canonical quantum gravity as a possible
framework to understand black holes. A different approach to
quantum gravity is superstring theory. It necessarily 
contains gravity and gauge theories, and must implement
supersymmetry for reasons of consistency.

 Like canonical quantum gravity, string theory
follows through the quantisation of a classical theory (a propagating
string in some background spacetime), but is itself interpreted
in a drastically different way:
It is supposed to give a fundamental theory where all interactions including
gravity are unified in a quantum framework. The background spacetime
used in the construction of the theory plays only an auxiliary
role. Like canonical quantum gravity, string theory suffers
from the ``problem of time", although this is not always
stated clearly. The notion of spacetime again
emerges only in an appropriate semiclassical limit.  
(The role of the semiclassical expansion parameter
is here played by the string length, see below).
An important fact in string theory is that consistency conditions (the absence
of a Weyl anomaly) severely {\em restricts} the number of dimensions
of this semiclassical spacetime, e.g. to $D=10$ for the superstring.
This, then, enforces the implementation of an appropriate
mechanism to encurl the superfluous dimensions in a Kaluza-Klein
type manner to avoid contradiction with observation.
Whether the level of canonical quantum gravity, as discussed in
Sect.~3, follows from string theory in an appropriate limit
is not yet clear. It must, however, lead to {\em some}
quantum gravitational corrections to the ordinary functional
Schr\"odinger equation, and may thus lead to the possibility
both to test the theory and to discriminate it from 
competitors like the approach presented in Sect.~3.

A detailed introduction into string theory can be found,
for example, in Polchinski (1994, 1996), and the references
therein. Here I only want to briefly sketch some intriguing recent
developments aiming at a derivation of the black hole entropy (3)
by counting quantum states, see Horowitz (1997) for a review.
String theory contains two important parameters: The string
length $l_S$ and the string coupling $g_S^2\equiv\exp(2\varphi)$.
Here, $\varphi$ denotes the dilaton field which appears in the
two-dimensional string action. It gives rise to the string coupling,
since $g_S^2$ appears as a ``gravitational constant"
 in the effective action (arising in the
semiclassical approximation to lowest order in $l_S$) for
the background spacetime and background fields. 
The Planck length, $l_P$, then appears as a {\em derived}
quantity,
 \be l_P\propto g_Sl_S \enspace , \ee
 and similar relations
follow for other ``coupling constants". 
It is important to note that the semiclassical approximation,
and with it the notion of a spacetime metric, breaks down
for curvatures bigger than $l_S^{-2}$.

How does the entropy of a black hole come into play?
First, assigning an entropy to an excited string state by counting
its degeneracy, it turns out that this entropy is (for high
excitations) proportional to the energy (mass) of that state and
not to the mass squared. It would thus seem as if a string had not
enough states to yield the entropy of a black hole.
The crucial point, however, is that the Planck length, and
therefore the gravitational constant, {\em depends} on the string coupling,
see (71). Thus, if $g_S$ is increased, $Gm$ is increased, too,
and a black hole is formed at some stage (Horowitz~1997).
Comparing, then, the black hole mass with the string mass
at $l_S=R_0$ ($R_0$ is the Schwarzschild radius), it turns out
that the black hole entropy becomes proportional to the string entropy.
A string may thus possess enough states to give the
Bekenstein-Hawking entropy.

For a quantitative comparison, one must give a precise
calculation. It is most straightforward in this respect to
first consider states which obey a relation similar to
$q=Gm$ in the Reissner-Nordstr\"om case
(although with generalised charges). Such states are called
BPS states.
 At weak coupling
($g_S\ll 1$), one has bound states of so-called D-branes
(Polchinski~1996) in flat space, and the number of these states
can be counted. D-branes are dynamical objects of various
dimensions, which are a necessary ingredient of string theory.
 As the coupling increases, the BPS-relation
between mass and charges is preserved, and the number of states
remains unchanged. For high coupling ($g_S\gg1$), one thus
obtains an extremal black hole with the same number of states.
Surprisingly, its Bekenstein-Hawking entropy exactly coincides
with the entropy of the D-branes in the flat space description
(Strominger and Vafa~1996). The original calculation was for
five-dimensional black holes and then generalised to four-dimensional
holes.
 One may thus interpret the D-branes
as giving the desired microscopic description for the black hole
entropy. Since it turns out in this approach that 
$S_{extreme}=A/4\hbar G\neq0$, string theory leads to a
different result than the canonical treatment presented in
Sect.~3.2. 

The calculations have been extended to the case of near-extremal
black holes which, in contrast to the extremal ones, exhibit
Hawking radiation (here interpreted as the emission of closed
strings from D-branes). It could be shown that even the {\em rate}
of Hawking radiation agrees with the decay amplitude for the
corresponding D-brane configuration
(see e.g. Das~1997). Since all string calculations preserve
unitarity, it seems that there is no violation of unitarity
also in the black hole radiation. Consequently, there
would be no ``loss of information". Of course, to get a
non-vanishing entropy in the first place, some coarse-graining
must be involved, and the process of decoherence will again 
play a crucial role (Myers~1997). There will thus only be the
apparent non-unitarity connected with the neglect of degrees of
freedom be present -- the total system evolves unitarily
(Giulini et al.~1996).  
 
Whether the above string result also holds for general black holes,
i.e. far away from extremality
(such as for the Schwarzschild black hole), is not yet clear.
It must also be emphasised that all results are obtained in
lowest order of $l_S$, i.e. in the lowest order of the semiclassical
approximation where a background structure is available. 
The full, non-perturbative, evolution of a black hole therefore
still remains mysterious.

In the semiclassical approximation to canonical quantum gravity,
as presented in Sect.~3, a crucial role for the interpretation
of entropy is played by the presence of boundary conditions
at the bifurcation sphere (where the two horizons in the
Kruskal diagramme meet). This, however, cannot be extended to the
full theory in a straightforward manner. The main reason is that
the horizon of a black hole is a {\em classical concept}. As I
emphasised in Sect.~3.1, the canonical theory does not possess
any notion of spacetime at the fundamental level, in the same way
as ordinary quantum theory does not possess any notion of particle
trajectories in the full theory. A horizon, however, is a genuine
spacetime concept. Therefore, the results presented in Sect.~3.2
only hold as far as a notion of spacetime can be applied at least
in some approximation. 

That the concept of an event horizon is a classical artifact,
becomes especially obvious in quantum cosmology (Zeh~1992).
Consider, for example, the case of a Friedmann universe that
classically recollapses. Since the entropy content of the present
universe is far from maximal, it must have been very tiny at
the big bang -- the big bang was extremely smooth (which is why
one would not expect to find many primordial black holes).
 This led
Penrose (1979) to the formulation of his Weyl tensor hypothesis
that the universe is homogeneous at the big bang, but not
at the big crunch. In quantum gravity, however, 
there is no external time parameter which could possibly 
distinguish between big bang and big crunch. If entropy is small
near the big bang, it must also be small near the big crunch,
since both regions correspond to the {\em same} region of the
quantum gravitational configuration space. The consequences of 
this fact for the arrow of time and for black holes were
investigated in Kiefer and Zeh (1995). Entropy is always growing
with increasing size of the Universe, leading to a (formal)
reversal of the arrow of time near the classical turning point.
The same boundary condition of low entropy at small size
necessarily leads to the fact that neither an event horizon
nor a singularity 
(naked or hidden) forms for a black hole.
 Cosmic censorship would thus be
automatically implemented. Although still speculative, this
scenario at least demonstrates what qualitatively new features
emerge from quantum gravity if one leaves the semiclassical sector.

\vskip 3mm
{\bf Acknowledgements}. I am grateful to Thorsten Brotz,
Valeri Frolov, Domenico
Giulini, and H.-Dieter Zeh for a critical reading of this
manuscript.
%
%
%

\end{document}